\documentclass{sf2a-conf2018}
\usepackage{graphicx}
\usepackage{hyperref}
\usepackage[]{natbib}  
\usepackage{epstopdf}

\def\BibTeX{{\rm B\kern-.05em{\sc i\kern-.025em b}\kern-.08em
    T\kern-.1667em\lower.7ex\hbox{E}\kern-.125emX}}
\bibpunct{(}{)}{;}{a}{}{,}  %%%%%%%%%%%%%  A&A bibliography style
\begin{document}

\TitreGlobal{SF2A 2018}

%%-----------------------------------------------------------------
%%      the top matter
%%

\title{The magnetic field of evolved hot stars}

\runningtitle{The magnetic field of evolved hot stars}

\author{C. Neiner}\address{LESIA, Paris Observatory, PSL University, CNRS, Sorbonne Université, Univ. Paris Diderot, Sorbonne Paris Cité, 5 place Jules Janssen, 92195 Meudon, France}

\author{A. Martin$^1$}
\author{G. Wade}\address{Department of Physics \& Space Science, Royal Military College of Canada, P.O. Box 17000, Station Forces, Kingston, Ontario, Canada, K7K 7B4}
\author{M. Oksala$^{1,}$}\address{Department of Physics, California Lutheran University, 60 West Olsen Road \#3700, Thousand Oaks, CA 91360, USA }

%% Keep this line, even if the page will be settled afterwards.
\setcounter{page}{237}

%%-----------------------------------------------------------------

\maketitle

%%-----------------------------------------------------------------
%%        The abstract
%% 
%%  Warning!  within the abstract:
%%  - do not use macros. 
%%  - do not use commands like: \cite, \citet, \citep ... etc.

\begin{abstract}
About 10\% of hot stars host a fossil magnetic field on the pre-main sequence and main sequence. However, the first magnetic evolved hot stars have been discovered only recently. An observing program has been set up to find more such objects. This will allow us to test how fossil fields evolve, and the impact of magnetism on stellar evolution. Already 7 evolved magnetic hot stars are now known and the rate of magnetic discoveries in the survey suggests that they host dynamo fields in addition to fossil fields. Finally, the weakness of the measured fields is compatible at first order with simple magnetic flux conservation, although the current statistics cannot exclude intrinsic decay or enhancement during stellar evolution.
\end{abstract}

%% Insert the keywords (to appear in the ADS indexing)
%% Keywords must be separated by a comma

%%-----------------------------------------------------------------

\section{Magnetism in hot stars}

On the pre-main sequence (PMS) and main sequence (MS), about 10\% of hot stars host a magnetic field detectable at their surface. It usually is mainly an oblique dipole field with a polar field strength between 300 G and 30 kG. These fields are intrinsically stable over decades. Only rotational modulation is recorded for various observables (the longitudinal field, UV lines sensitive to the magnetised wind, H$\alpha$ and X-ray emission coming from the magnetosphere,...), due to the obliquity between the magnetic axis and the rotation axis. Analytical work, numerical simulations, and observations converge to say that the origin of these fields is fossil, i.e. they are the remnants of a magnetic field already present in the molecular cloud during star formation, possibly enhanced by a dynamo action during the early phases of the life of the star. More details can be found about the observations and origin of magnetic fields in hot stars in the review papers by \cite{grunhut2015} and \cite{neiner2015}, respectively. 

How these fossil fields evolve past the main sequence, whether magnetic flux is conserved or whether there is magnetic decay or enhancement, is not yet known. Moreover, the presence of such magnetic fields must have a strong impact on stellar evolution, because of the interaction of the field with interior fluid motions, impacting the internal rotation profile, angular momentum, and chemical transport. In addition, the field also interacts with the stellar wind producing magnetic braking and a reduction of mass loss. Finally, the magnetic field contributes to the energy budget of supernova explosions and thus directly impacts the stellar death and the resultant remnants.

Furthermore, as a hot star evolves, convective regions appear in the radiative envelope. Dynamo fields could develop and interact with the initial fossil field. 3D MHD simulations \citep{featherstone2009} show that such interaction would increase the strength of the dynamo field and modify the obliquity of the fossil field. 

\section{Magnetic flux conservation vs decay/enhancement}

Magnetic flux conservation is the simplest theory that one can consider for the evolution of the magnetic field in hot stars. It assumes that nothing intrinsically weakens or strengthens the field. However, as the star evolves, its radius increases, and therefore the field measured at the stellar surface decreases with the following law: $B(t_2) = B(t_1) \frac{R(t_1)^2}{R(t_2)^2}$, where $B$ is the field strength and $R$ is the radius at time $t_1$ and $t_2$. In other words, when the radius increases by a factor 10, the surface field decreases by a factor 100. Since hot stars typically have polar fields of 3 kG on the MS, one expects from magnetic flux conservation that field strengths at the surface of hot supergiants are of the order of 5-10 G. These predictions are in good agreement with the general lack of detected fields in evolved hot stars in recent spectropolarimetric surveys, such as MiMeS \citep{wade2016} or BOB \citep{morel2014}, for which the detection limit was typically 100 G. 

The evolution of surface magnetic fields in OBA stars during the MS has been investigated by \cite{bagnulo2006}, \cite{landstreet2007,landstreet2008}, and more recently by \cite{fossati2016}. These studies provide convincing evidence that the strengths of surface magnetic fields decrease systematically during the MS, in response to stellar expansion but also possibly to Ohmic decay and other (currently unknown) mechanisms. Therefore, it is possible that an additional process intrinsically decreases the field strength of evolved hot stars in addition to the apparent surface field decrease. However, the aforementioned studies could not investigate what happens to the magnetic fields after the MS and the exact processes leading to the field decrease during evolution until their late stellar stages. Only the direct spectropolarimetric study of magnetic evolved hot stars can help answer these questions.

\section{Known magnetic evolved hot stars}

\begin{figure}[t!]
 \centering
 \includegraphics[width=\textwidth,clip]{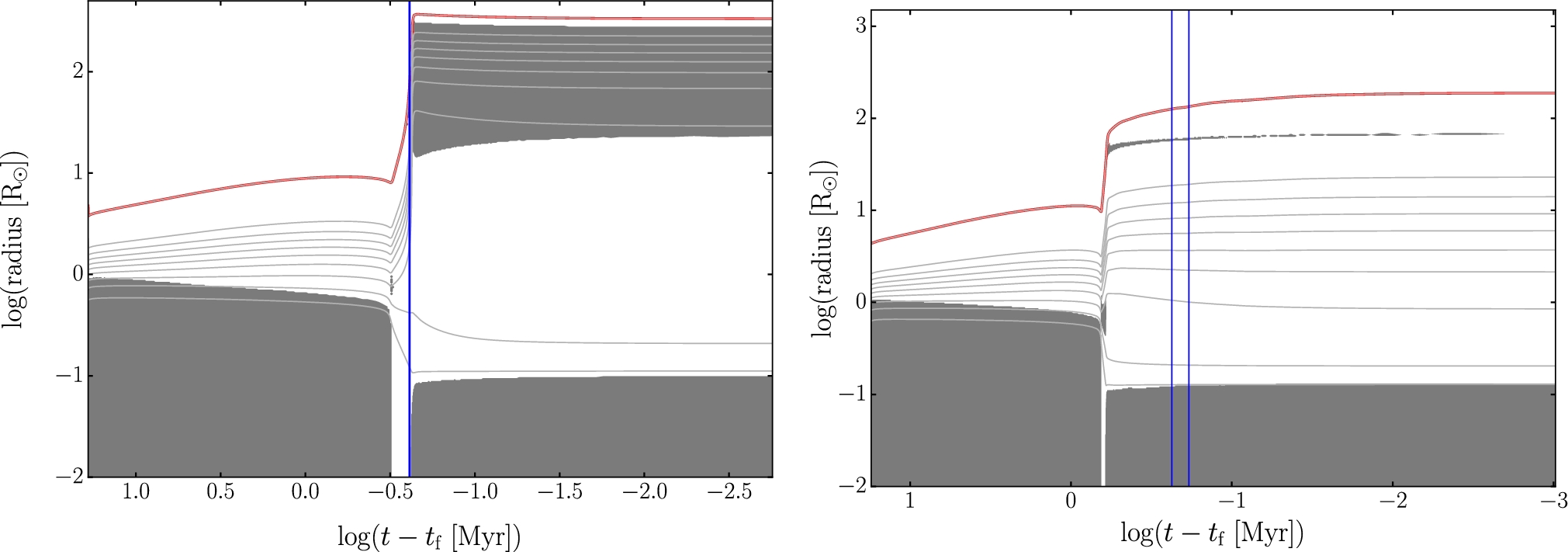}      
  \caption{Kippenhahn diagram showing the evolution of the stellar structure of $\iota$\,Car (left) and HR\,3890 (right). Convective zones are indicated in grey, while radiative zones are in white. The red line indicates the stellar surface, while the vertical blue lines show the current position of the star in this diagram, within uncertainties. From \cite{neiner2017}.}
  \label{kippenhahn}
\end{figure}

Until recently only a few magnetic evolved hot stars were known. The O9.5I star $\zeta$\,Ori\,Aa was discovered to be magnetic by \cite{bouret2008} but the presence of its field was confirmed only recently \citep{blazere2015}. It has a polar field strength of $\sim$140 G. However, \cite{blazere2015} showed that the star is located just past the MS in the HR diagram with a radius of about 20 R$_\odot$. $\epsilon$\,CMa is a B1.5II star discovered to be magnetic by \cite{fossati2015}, but its evolutionary status was unclear. A new spectropolarimetric study by \cite{neiner2017} confirmed the presence of a field with a polar strength of about 35 G but showed that the star is located at the end of the MS. Eventually, two well evolved hot stars were discovered in 2017 \citep{neiner2017} thanks to the BritePol survey (which observed with spectropolarimetry all stars brighter than V=4, see \cite{neiner2014}). $\iota$\,Car is an A7Ib supergiant with a 3 G polar field strength and a radius of approximately 60 R$_\odot$, which is either on its first crossing of the HR diagram or on the blue loop. Unfortunately $\iota$\,Car is just at the transition between a fully radiative envelope and the creation of an external convective zone (see Fig.~\ref{kippenhahn}, left), therefore it is not an easy target to study evolutionary effects. HR\,3890 is an A8Ib supergiant with a 6 G polar field strength and a radius above 100 R$_\odot$, which is on its first and only crossing of the HR diagram. HR\,3890 has already developed an external convective region but it is very thin (see Fig.~\ref{kippenhahn}, right) and likely only has a weak impact on the fossil magnetic field. See \cite{neiner2017} for more details on $\iota$\,Car and HR\,3890. As a consequence, until recently, no appropriate target was known to carefully study the effect of stellar evolution on a fossil magnetic field, nor the evolution of the field itself.

\section{The LIFE project}

\begin{figure}[t!]
 \centering
 \includegraphics[width=0.48\textwidth,clip]{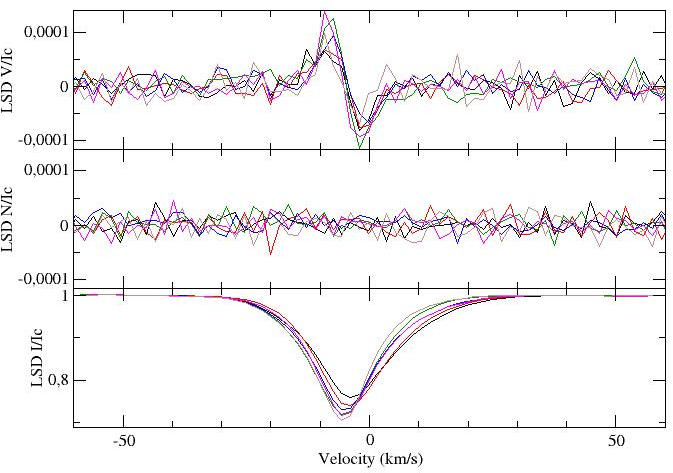}      
 \includegraphics[width=0.48\textwidth,clip]{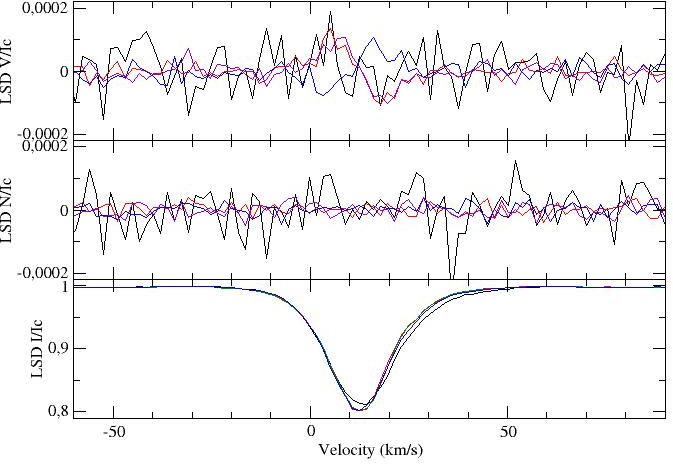}\\      
 \includegraphics[width=0.48\textwidth,clip]{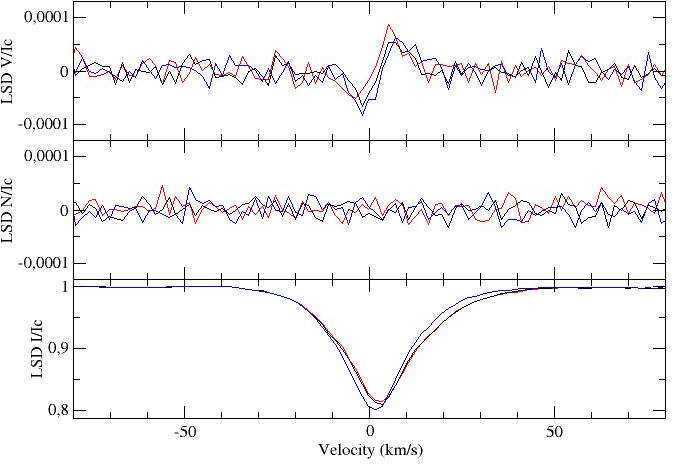}      
 \includegraphics[width=0.48\textwidth,clip]{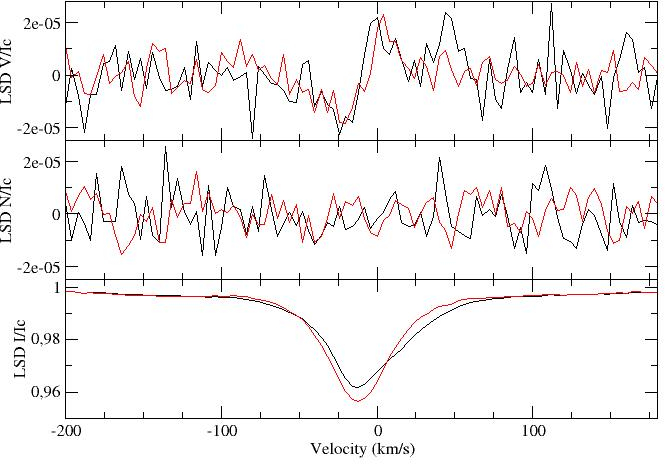}      
  \caption{LSD Stokes V (top), Null polarisation (middle), and Stokes I (bottom) profiles of 19\,Aur observed with ESPaDonS between September 2016 and October 2018 (top left), 13\,Mon observed with ESPaDonS between February 2016 and January 2018 (top right), $\eta$\,Leo observed with ESPaDonS in November 2017 and January 2018 (bottom left), and d\,Car observed with HarpsPol in February 2018 (bottom right).}
  \label{19aur}
\end{figure}

The LIFE (the Large Impact of magnetic Fields on the Evolution of hot stars) was set up to discover and study more magnetic evolved hot stars thanks to  spectropolarimetric observing programs with ESPaDOnS at CFHT and HarpsPol at ESO. The goal is to search for magnetic fields in class I, II, and III, OBA stars with a spectropolarimetric measurement precision better than 1 G. The observations started in 2016 and have already led to the publication of two magnetic detections \citep{martin2018}. 19\, Aur is a A5Ib star with a 3 G polar field strength and a radius of 40-50 R$\odot$. However, its magnetic signature does not vary much from one observation to the next (see Fig.~\ref{19aur}, top left), indicating either that it is in a particular geometrical configuration or that the rotation period of the star is very long. In both cases, a full characterisation of the field will not be easy. HR\,3042 is a B8/9II star with a 760 G polar field and a rotation period of about 1 week. However, the spectrum and field of this target are rather reminiscent of MS hot stars and it is likely that this star was misclassified as evolved in the literature. Since the \cite{martin2018} paper, 5 more magnetic detections have been obtained within the LIFE project. HD\,167686 is a B8II star with a 1 kG polar field and HR\,3867 is a B9IIpSi star with a polar field of 600 G. There again, one may wonder if these two stars were properly classified; they may rather be MS hot stars. On the other hand, $\eta$\,Leo and 13 Mon are A0I supergiants that are clearly evolved and have a polar field strength of 3 G and 9 G, respectively (see Fig.~\ref{19aur}, top right and bottom left). Finally, d\,Car is a B1III star with a polar field strength of 6 G (see Fig.~\ref{19aur}, bottom right). The latter 3 targets are excellent candidates for a follow-up spectropolarimetric study to fully characterise their field. Such observations are already ongoing for $\eta$\,Leo and 13 Mon.

\section{Conclusions}

A few dozens of evolved hot stars have already been observed with very high-precision spectropolarimetry. Some of them turned out to be less evolved than claimed in the literature. Despite this, at least 7 evolved magnetic hot stars are now identified. Statistics are still sparse, nevertheless the large number of magnetic detections already obtained seems to indicate that the magnetic fraction in evolved hot stars is higher than in PMS and MS stars, in particular in luminosity class I objects. The increased magnetic detection rate indicates that dynamos may well be observed in addition to the fossil fields in evolved stars, likely due to the appearance of one or more external convection zones in the radiative enveloppe. A detailed study of the newly detected magnetic evolved hot stars will allow us to better characterize the observed fields. 

Moreover, the very weak field strength measured at the surface of these magnetic evolved hot stars is compatible with the strengths expected from simple magnetic flux conservation at first order, although they do not exclude possible decays or enhancements (including dynamos). Future new magnetic detections from the LIFE survey will allow us to improve the statistical results and better test the various scenarios of magnetic field evolution.

\begin{acknowledgements}
This work as made used of the SIMBAD database operated at CDS, Strasbourg (France), and of NASA's Astrophysics Data System (ADS). We thank the Paris Observatory for its financial support. 
\end{acknowledgements}

%% The following lines are required when using BibTEX (strongly encouraged!):
\bibliographystyle{aa}  % A&A bibliography style file (aa.bst)
\bibliography{sf2a-Neiner-MagEvol} % your references in file: Yourfile.bib

\begin{thebibliography}{15}
\expandafter\ifx\csname natexlab\endcsname\relax\def\natexlab#1{#1}\fi

\bibitem[{{Bagnulo} {et~al.}(2006){Bagnulo}, {Landstreet}, {Mason}, {Andretta},
  {Silaj}, \& {Wade}}]{bagnulo2006}
{Bagnulo}, S., {Landstreet}, J.~D., {Mason}, E., {et~al.} 2006, \aap, 450, 777

\bibitem[{{Blaz{\`e}re} {et~al.}(2015){Blaz{\`e}re}, {Neiner}, {Tkachenko},
  {Bouret}, \& {Rivinius}}]{blazere2015}
{Blaz{\`e}re}, A., {Neiner}, C., {Tkachenko}, A., {Bouret}, J.-C., \&
  {Rivinius}, T. 2015, \aap, 582, A110

\bibitem[{{Bouret} {et~al.}(2008){Bouret}, {Donati}, {Martins}, {Escolano},
  {Marcolino}, {Lanz}, \& {Howarth}}]{bouret2008}
{Bouret}, J.-C., {Donati}, J.-F., {Martins}, F., {et~al.} 2008, \mnras, 389, 75

\bibitem[{{Featherstone} {et~al.}(2009){Featherstone}, {Browning}, {Brun}, \&
  {Toomre}}]{featherstone2009}
{Featherstone}, N.~A., {Browning}, M.~K., {Brun}, A.~S., \& {Toomre}, J. 2009,
  \apj, 705, 1000

\bibitem[{{Fossati} {et~al.}(2015){Fossati}, {Castro}, {Morel}, {Langer},
  {Briquet}, {Carroll}, {Hubrig}, {Nieva}, {Oskinova}, {Przybilla},
  {Schneider}, {Sch{\"o}ller}, {Sim{\'o}n-D{\'{\i}}az}, {Ilyin}, {de Koter},
  {Reisenegger}, \& {Sana}}]{fossati2015}
{Fossati}, L., {Castro}, N., {Morel}, T., {et~al.} 2015, \aap, 574, A20

\bibitem[{{Fossati} {et~al.}(2016){Fossati}, {Schneider}, {Castro}, {Langer},
  {Sim{\'o}n-D{\'{\i}}az}, {M{\"u}ller}, {de Koter}, {Morel}, {Petit}, {Sana},
  \& {Wade}}]{fossati2016}
{Fossati}, L., {Schneider}, F.~R.~N., {Castro}, N., {et~al.} 2016, \aap, 592,
  A84

\bibitem[{{Grunhut} \& {Neiner}(2015)}]{grunhut2015}
{Grunhut}, J.~H. \& {Neiner}, C. 2015, in IAU Symposium, Vol. 305, Polarimetry,
  ed. K.~N. {Nagendra}, S.~{Bagnulo}, R.~{Centeno}, \& M.~{Jes{\'u} s
  Mart{\'{\i}}nez Gonz{\'a}lez}, 53--60

\bibitem[{{Landstreet} {et~al.}(2007){Landstreet}, {Bagnulo}, {Andretta},
  {Fossati}, {Mason}, {Silaj}, \& {Wade}}]{landstreet2007}
{Landstreet}, J.~D., {Bagnulo}, S., {Andretta}, V., {et~al.} 2007, \aap, 470,
  685

\bibitem[{{Landstreet} {et~al.}(2008){Landstreet}, {Silaj}, {Andretta},
  {Bagnulo}, {Berdyugina}, {Donati}, {Fossati}, {Petit}, {Silvester}, \&
  {Wade}}]{landstreet2008}
{Landstreet}, J.~D., {Silaj}, J., {Andretta}, V., {et~al.} 2008, \aap, 481, 465

\bibitem[{{Martin} {et~al.}(2018){Martin}, {Neiner}, {Oksala}, {Wade},
  {Keszthelyi}, {Fossati}, {Marcolino}, {Mathis}, \& {Georgy}}]{martin2018}
{Martin}, A.~J., {Neiner}, C., {Oksala}, M.~E., {et~al.} 2018, \mnras, 475,
  1521

\bibitem[{{Morel} {et~al.}(2014){Morel}, {Castro}, {Fossati}, {Hubrig},
  {Langer}, {Przybilla}, {Sch{\"o}ller}, {Carroll}, {Ilyin}, {Irrgang},
  {Oskinova}, {Schneider}, {D{\'{\i}}az}, {Briquet}, {Gonz{\'a}lez},
  {Kharchenko}, {Nieva}, {Scholz}, {de Koter}, {Hamann}, {Herrero},
  {Ma{\'{\i}}z Apell{\'a}niz}, {Sana}, {Arlt}, {Barb{\'a}}, {Dufton},
  {Kholtygin}, {Mathys}, {Piskunov}, {Reisenegger}, {Spruit}, \&
  {Yoon}}]{morel2014}
{Morel}, T., {Castro}, N., {Fossati}, L., {et~al.} 2014, The Messenger, 157, 27

\bibitem[{{Neiner} \& {L{\`e}bre}(2014)}]{neiner2014}
{Neiner}, C. \& {L{\`e}bre}, A. 2014, in SF2A-2014: Proceedings of the Annual
  meeting of the French Society of Astronomy and Astrophysics, ed. J.~{Ballet},
  F.~{Martins}, F.~{Bournaud}, R.~{Monier}, \& C.~{Reyl{\'e}}, 505--508

\bibitem[{{Neiner} {et~al.}(2015){Neiner}, {Mathis}, {Alecian}, {Emeriau},
  {BinaMIcS}, \& {MiMeS Collaborations}}]{neiner2015}
{Neiner}, C., {Mathis}, S., {Alecian}, E., {et~al.} 2015, in IAU Symposium,
  Vol. 305, Polarimetry, ed. K.~N. {Nagendra}, S.~{Bagnulo}, R.~{Centeno}, \&
  M.~{Jes{\'u} s Mart{\'{\i}}nez Gonz{\'a}lez}, 61--66

\bibitem[{{Neiner} {et~al.}(2017){Neiner}, {Oksala}, {Georgy}, {Przybilla},
  {Mathis}, {Wade}, {Kondrak}, {Fossati}, {Blaz{\`e}re}, {Buysschaert}, \&
  {Grunhut}}]{neiner2017}
{Neiner}, C., {Oksala}, M.~E., {Georgy}, C., {et~al.} 2017, \mnras, 471, 1926

\bibitem[{{Wade} {et~al.}(2016){Wade}, {Neiner}, {Alecian}, {Grunhut}, {Petit},
  {Batz}, {Bohlender}, {Cohen}, {Henrichs}, {Kochukhov}, {Landstreet},
  {Manset}, {Martins}, {Mathis}, {Oksala}, {Owocki}, {Rivinius}, {Shultz},
  {Sundqvist}, {Townsend}, {ud-Doula}, {Bouret}, {Braithwaite}, {Briquet},
  {Carciofi}, {David-Uraz}, {Folsom}, {Fullerton}, {Leroy}, {Marcolino},
  {Moffat}, {Naz{\'e}}, {Louis}, {Auri{\`e}re}, {Bagnulo}, {Bailey},
  {Barb{\'a}}, {Blaz{\`e}re}, {B{\"o}hm}, {Catala}, {Donati}, {Ferrario},
  {Harrington}, {Howarth}, {Ignace}, {Kaper}, {L{\"u}ftinger}, {Prinja},
  {Vink}, {Weiss}, \& {Yakunin}}]{wade2016}
{Wade}, G.~A., {Neiner}, C., {Alecian}, E., {et~al.} 2016, \mnras, 456, 2

\end{thebibliography}

\end{document}